\documentclass[twocolumn,showpacs,amsmath,amssymb,aps,prl,floatfix,nofootinbib]{revtex4}
\usepackage{graphicx}
\usepackage{dcolumn}
\usepackage{bm}
\usepackage{graphicx}
\usepackage{amsmath}

\begin{document}


\title{Enhancing efficacy in laser projection by four wavelength combination}


\author{Ian Wallhead, Roberto Oca\~{n}a$^*$ and Paula Quinz\'{a}}

\affiliation{AIDO, Instituto Tecnol\'{o}gico de \'{O}ptica,
C/Nicol\'{a}s Cop\'{e}rnico, 7-13, Parque Tecnol\'{o}gico, 46980
Paterna, Valencia, Spain
\\
$^*$Corresponding author: rocana@aido.es, robertoocagna@gmail.com}

\begin{abstract}
In this letter we present a novel study about the combination of
four laser wavelengths in order to maximize both color gamut and
efficacy to produce the color white. Firstly, an analytic method
to calculate efficacy as function of both four laser wavelengths
and four laser powers is derived. Secondly we provide a new way to
present the results by providing the diagram efficacy vs color
gamut area that summarizes the performance of any wavelength
combination for projection purposes. The results indicate that the
maximal efficacy for the D65 white is only achievable by using a
suitable combination of both laser power ratios and wavelengths.
\end{abstract}
\maketitle



Mobile projection devices (usually known as picoprojectors) should
be designed to maximize their luminous efficacy. 
This is for two main reasons.  Firstly, being either stand alone
devices or embedded in other products, they are likely to be
powered by battery, and lifetime an important factor~\cite{Urey}.
Secondly, the increasing use of lasers to project images calls for
a consideration of eye safety issues~\cite{SWallhead}. The
brightness of the projected image may be limited by the Class II
accessible emission limit. There is reason to believe that current
laser beam scanning picoprojector technology is already close to
the power ceiling based on eye safety limits~\cite{SBuckley}.
Consequently, it would be desirable to improve luminous efficacy
to increase the output luminous flux for the same eye-safe optical
power limit.

The relationship between the choice of laser wavelengths and the
resulting luminous efficacy has been discussed in the literature
\cite{EBWavelegths,SWallhead2,Patente}. Here we present the
opportunity to increase luminous efficacy by adding a fourth laser
wavelength. Usually, the three primary wavelengths of a laser
display offer a wide color gamut and so the emphasis in this study
has been to maximize the safe luminous flux whilst maintaining the
gamut, rather than increasing the gamut. Utilization of additional
wavelengths or colors in display and illumination applications is
however not new.  Reasons for adding one or more additional colors
have traditionally been for improved gamut and/or improved
brightness and have usually applied to liquid crystal display
applications. Brown et al. \cite{Brown} have discussed adding a
white pixel to the red, green and blue pixels of an LCD display to
increase the brightness of a display. Sharp has recently launched
a four-color television with claims of increased brightness and
color gamut~\cite{sharp}.

Here we present a study of the choices of four laser wavelengths
to optimize luminous efficacy for the projection of the CIE
standard illuminant D65 (daylight white). Since this is a display
application, the basis for color mixing calculations is the set of
1964, $10^{\circ}$ CIE color matching functions~\cite{CIE}. The
CIE recommends the use of the color-matching functions of the CIE
1964 Supplementary Standard Colorimetric Observer ``whenever
correlation with visual color matching of fields of angular
subtense greater than about $4^{\circ}$ at the eye of the observer
is desired''. It has also been noted in the literature that in the
optimization of the efficacy of a 3-wavelength projector there is
a trade-off with the color gamut~\cite{SWallhead}.  For this
reason, in this study we compare the efficacy with the color gamut
area as presented on the CIE 1976 UCS (uniform chromaticity scale)
diagram.

The tristimulus values (X, Y and Z) are determined from the color
matching functions, $\overline{x}(\lambda)$,
$\overline{y}(\lambda)$, and $\overline{z}(\lambda)$, of a given
illuminant spectral power distribution $P(\lambda)$ as follows:
\begin{align}
G_n=\int^{\infty}_{0}P(\lambda)\overline{g}_{n}(\lambda)d\lambda ,
\end{align}
where $G_n$ and $\overline{g}_{n}(\lambda)$ represent the
sequences $ (X,Y,Z)$ and $
(\overline{x}(\lambda),\overline{y}(\lambda),\overline{z}(\lambda))$
respectively. For the case of a three wavelength combination if we
define a matrix M as
\begin{align}
\textbf{M}=\left[
\begin{array}{lll}
\overline{x}(\lambda_{r})& \overline{x}(\lambda_{g})& \overline{x}(\lambda_{b})\\
\overline{y}(\lambda_{r}) & \overline{y}(\lambda_{g}) & \overline{y}(\lambda_{b})\\
\overline{z}(\lambda_{r}) & \overline{z}(\lambda_{g}) &
\overline{z}(\lambda_{b})
\end{array}\right],
\end{align}
the power of red, green and blue lasers can be calculated by
\begin{align}\label{eq1}
\textbf{P}_{rgb}(\boldsymbol{\lambda})=\textbf{M}^{-1}\textbf{W} ,
\end{align}
where the vector $\textbf{W}$ contains the $x$, $y$ and $z$
chromaticity coordinates of white D65 point, and
$\boldsymbol{\lambda}$ the wavelengths $\lambda_r$, $\lambda_g$,
and $\lambda_b$. In the framework of the CIE 1964 $10^\circ$
color-matching functions, the coordinates x, y and z are 0.31382,
0.331 and 0.35518 respectively. $\textbf{P}_{rgb}$ describes the
unique solution of the combination of relative powers of the red,
green and blue wavelengths to produce D65 white. For a
four-wavelength combination there is no single solution. The
combination of four colors is dependent on the relative power of
the fourth wavelength.  As will be demonstrated this fourth
wavelength is consistently found to be in the yellow region and so
is represented as $\lambda_y$.  To calculate the four color
combination we first define a new vector $\textbf{\~{W}}$:

\begin{align}
\textbf{\~{W}}=\left[
\begin{array}{l}
x_w-P_y \overline{x}(\lambda_y)\\
y_w-P_y \overline{y}(\lambda_y)\\
z_w-P_y \overline{z}(\lambda_y)
\end{array}\right]
\end{align}

Expression~\ref{eq1} can be now rewritten in terms of
$\textbf{\~{W}}$ for the case of a four wavelength combination as
follows:
\begin{align}\label{eq2}
\textbf{P}_{rgb}(\boldsymbol{\lambda}, \lambda_y,
P_y)=\textbf{M}^{-1}\textbf{\~{W}}
\end{align}
In the former expression, $\textbf{{P}}_{rgb}$ depends on $P_y$
and $\lambda_y$ as well. If we redefine $\boldsymbol{\lambda}$ as
the four-dimensional vector [$\lambda_r$, $\lambda_g$,
$\lambda_b$, $\lambda_y$], the power vector containing the
contribution of all laser sources might be written as follows:

\begin{align}
\textbf{P}=\left[
\begin{array}{llll}
P_r(\boldsymbol{\lambda},P_y),& P_g(\boldsymbol{\lambda},P_y),&
P_b(\boldsymbol{\lambda},P_y),& P_y
\end{array}\right]
\end{align}

Since the combination of colors is in terms of relative powers we
arbitrarily choose to define the powers of each wavelength with
respect to the blue wavelength.  Hence, we define a new power
vector $\textbf{\~{P}}$ as $\textbf{P}/P_b$. Thus, if values for
the components red, green, blue and yellow of the vector
$\boldsymbol{\lambda}$ and the ratio $P_y /P_b$ are selected,
expression~\ref{eq2} provides a unique result for the ratios
$P_r/P_b$ and $P_g/P_b$. The total power normalized by $P_b$ is
\begin{align}
\textmd{\~{P}}_t=\frac{P_r}{P_b}+\frac{P_g}{P_b}+1+\frac{P_y}{P_b}
\end{align}
In the same way, using the values ${P_r}/{P_b}$, ${P_g}/{P_b}$ and
$P_y/P_b$, a normalized flux $\textmd{\~{L}}_t$ can be obtained.
So, the efficacy is
\begin{align}
E_{ff}=\frac{\textmd{\~{L}}_t}{\textmd{\~{P}}_t}=\frac{L_t}{P_t}
\end{align}
This method allows us to calculate the efficacy by choosing the
relative power of the yellow source with respect to the blue
source.

Since the CIE 1976 chromaticity diagram is known to be
perceptively relatively linear~\cite{Ian2} it is appropriate to
quantify the extent of the color gamut as the enclosed area as
represented on this plot~\cite{EBWavelegths}. In doing so any
trade-off between efficacy and gamut can be evaluated. To do this,
we firstly have to calculate the x, y and z coordinates for each
laser wavelength. These coordinates can be obtained from the
following expression:
\begin{align}\label{eq3}
g_n=\frac{G_n}{\sum_{n=1}^3 G_n} ,
\end{align}
where $g_{n}$ represent one element of the sequence $(x,y,z)$.
From this set of coordinates, the $u'$ and $v'$ of the CIE 1976
chart are obtained as follows:

\begin{align}
u'=\frac{4x}{3+12y-2x}
\end{align}

\begin{align}
v'=\frac{9y}{3+12y-2x}
\end{align}

\begin{figure}[htb]
\centerline{\includegraphics[width=\columnwidth]{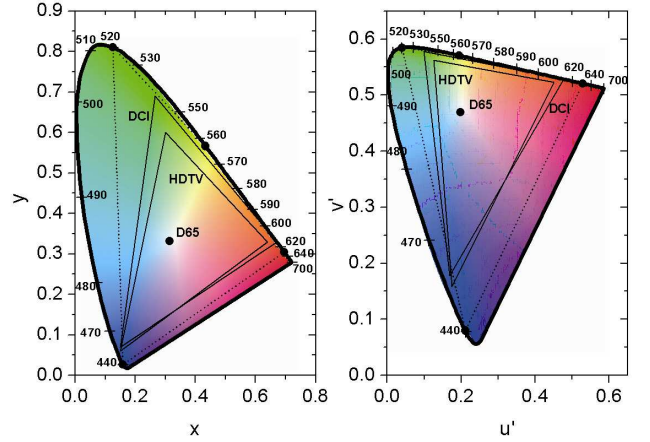}}
\caption{
Left part: CIE 1931 chart. Right part: CIE
1976 chart. In both graphs, achievable color gamuts  by the DCI
and HDTV input signals have been plotted. Dashed line corresponds
to the optimal solution found here using the laser wavelengths
630~nm, 520~nm, 441~nm and 563~nm}\label{F2}
\end{figure}

The color gamut area can be considered as a triangle formed by the
coordinates ($u'_r$,$v'_r$), ($u'_g$,$v'_g$) and ($u'_b$,$v'_b$)
of the red, green and blue lasers respectively if $\lambda_g \geq
520~nm$. As fig.~\ref{F2} shows, the yellow coordinates are in
this case almost on the line formed by the green and red
coordinates and therefore the contribution of the yellow laser to
the formation of a more complex area can be neglected for more
simplicity. Thus, under this approximation the color gamut area
(A) is:
\begin{align}\label{CGA}
A = \frac{1}{2}[u'_b(v'_r - v'_g)+ u'_g(v'_b - v'_r)+ u'_r(v'_g -
v'_b)]
\end{align}

In order to obtain solutions that could be of practical interest
and that can be compared with current display devices, we have
selected wavelength ranges that try to cover the most important
signal inputs, i.e. DCI and HDTV signal inputs plotted in
fig.~\ref{F2}~\cite{HDTV,DCI}. In this way, a laser projector
could display the color spectrum managed by these inputs. The
selected wavelength ranges are as follows: $600~nm\leq
\lambda_r\leq 630~nm$, $520~nm\leq \lambda_g\leq 550~nm$,
$440~nm\leq \lambda_b\leq 470~nm$ and $560~nm\leq \lambda_y\leq
590~nm$.

\begin{figure}[htb]
\centerline{\includegraphics[width=\columnwidth]{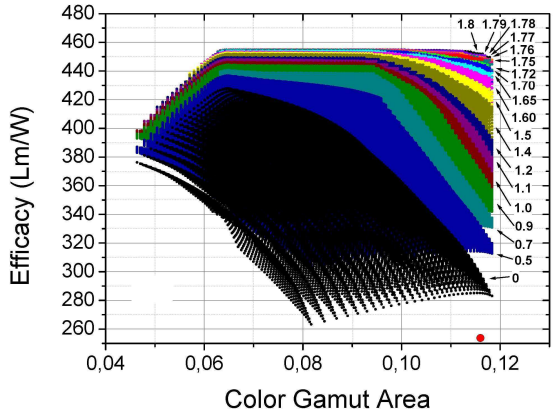}}
\caption{
Efficacy of the white point D65 as
function of the color gamut area for different ratios of the
yellow laser power to the blue laser power. The big isolated
circle denotes the commercial projector described in the
text.}\label{F1}
\end{figure}

Fig.~\ref{F1} contains the positive solutions of over 21 million
calculations of equation (\ref{eq2}) at the above specified
wavelength ranges and several power ratios ${P_y}/{P_b}$. The
efficacy-gamut trade-off of a three color mix is clear.  As the
power of the fourth wavelength is increased the efficacy of the
white point D65 increases at the wavelengths studied. If we choose
a figure of merit of the product of efficacy and color gamut area,
the most optimum value occurs at ${P_y}/{P_b}= 1.8$, ${P_r}/{P_b}=
9\times 10^{-3}$ and ${P_g}/{P_b}= 7\times 10^{-3}$ and at the
wavelengths 630~nm of the red laser, 520~nm of the green laser,
441~nm of the blue laser and 563~nm of the yellow laser which
produce an efficacy of 449.7~$lm/W$. Since the power needed for
the formation of D65 white with the red and green laser is
negligible, the calculation results indicate that this choice for
both optimum efficacy and color gamut area is mainly obtained with
the contributions of the blue and yellow laser. In the $u'$, $v'$
space this means that the chromaticity coordinate of the D65 white
point is on the straight line formed by the blue laser and yellow
laser coordinates. Note that the red and green wavelengths can be
considered in the developed method as free variables as well and
therefore are obtained as the best optimum wavelengths which
together with the blue wavelength maximize the color gamut area in
expression~\ref{CGA}.

By comparison, a commercially available three-color picoprojector
(Microvision Show WX~\cite{micro}) with wavelengths of $442~nm$,
$532~nm$ and $642~nm$ has an efficacy of $254~{lm}/W$. The
performance of this device has been plotted in fig.~\ref{F1} by
means of the coordinates of the efficacy and color gamut area. An
addition of a yellow laser at $564~nm$ ($Py = 1.76 Pb$) would
increase the efficacy by $77\%$ to $451~{lm}/W$.

One additional point is worth mentioning. Fig.~\ref{F1} shows that
there is a very distinct upper limit to the efficacy of any color
mix. This occurs at $\sim 455~lm/W$.  Insofar as the photopic
curve presents the maximum efficacy for a single wavelength
(683~lm/W at 555~nm), 455~lm/W has to be understood to be the
maximum efficacy with which the color white (D65 point) can be
projected by any combination of illumination sources.

The shape of the areas shown in fig.~\ref{F1} depends on the
wavelength ranges used. In fact, if we decrease the color gamut
area by using other wavelength ranges closer to the D65 white
point in the $u'$,$v'$ space, the efficacy would increase up to
the maximum value of $\sim 455~lm/W$ but at the cost of the color
gamut area limiting the capability for obtaining different colors.

In conclusion, we have derived a formulation that permits to
calculate the best combination of four wavelengths and powers for
laser projection maximizing both color gamut area and efficacy of
the D65 white point. The results show that the maximal efficacy
projected using the most common video input signals can be only
achieved by adding a fourth laser wavelength. Combination of three
laser wavelengths would produce less efficacy at suitable color
gamuts for projection. The best selection of wavelengths and
powers in this framework is found at the wavelengths 630~nm,
520~nm, 441~nm and 563~nm for the red, green, blue and yellow
lasers and the power ratio ${P_y}/{P_b}= 1.8$ that would produce
an efficacy of 449.7~$lm/W$ when forming a D65 white. These
conclusions and results represent a guideline for the future
development of picoprojectors and cinema projectors in which
particular emphasis is placed on the use of high laser power to
increase luminous flux and its implications in safety regulations.

\pagebreak

\end{document}